%% file: main.tex
\documentclass[conference]{new-aiaa}

\usepackage[utf8]{inputenc}
\usepackage{textcomp}

\usepackage{graphicx}
\usepackage{amsmath,bm}
\usepackage[version=4]{mhchem}
\usepackage{siunitx}
\usepackage{longtable,tabularx}
\usepackage[nolist]{acronym}
\usepackage{todonotes}
\usepackage{mathtools}

\author{Rasmus Steffensen\footnote{Research Associate, Institute of Flight System Dynamics, Bolzmann Strasse 15, 85748 Garching bei München, Germany.} and Agnes Steinert\footnote{Research Associate, Institute of Flight System Dynamics, Bolzmann Strasse 15, 85748 Garching bei München, Germany.}}
\affil{Technical University of Munich, Garching, 85748, Germany}
\author{Ewoud J.J. Smeur\footnote{Assistant Professor, Control and Simulation, Kluyverweg 1, 2629HS Delft, The Netherlands.}}
\affil{Delft University of Technology, Delft, Zuid-Holland, 2629HS, The Netherlands}

\title{Nonlinear Dynamic Inversion with Actuator Dynamics: an Incremental Control Perspective
\footnotetext{All authors contributed equally to the contributions of this paper}}

\begin{document}
\begin{acronym}
\acro{INDI}{Incremental Nonlinear Dynamic Inversion}
\acro{NDI}{Nonlinear Dynamic Inversion}
\acro{ANDI}{Actuator-NDI}
\end{acronym}

\maketitle

\begin{abstract}
In this paper, we derive a sensor based \ac{NDI} control law for a nonlinear system with first-order linear actuators, and compare it to \ac{INDI}, which has gained popularity in recent years.
It is shown that for first-order actuator dynamics, \ac{INDI} approximates the corresponding \ac{NDI} control law arbitrarily well under the condition of sufficiently fast actuators.
If the actuator bandwidth is low compared to changes in the states, the derived \ac{NDI} control law has the following advantages compared to \ac{INDI}: 1) compensation of state derivative terms 2) well defined error dynamics and 3) exact tracking of a reference model, independent of error controller gains in nominal conditions.
The comparison of the \ac{INDI} control law with the well established control design method \ac{NDI} adds to the understanding of incremental control. It is additionally shown how to quantify the deficiency of the \ac{INDI} control law with respect to the exact \ac{NDI} law for actuators with finite bandwidth.
The results are confirmed through simulation results of the rolling motion of a fixed wing aircraft.
\end{abstract}


\acresetall

\section*{Nomenclature}


{\renewcommand\arraystretch{1.0}
\noindent\begin{longtable*}{@{}l @{\quad=\quad} l@{}}
$\boldsymbol{x}$  & State vector \\
$\boldsymbol{u}$  & Input vector \\
$\boldsymbol{y}$  & Output vector \\
$\boldsymbol{y}_{ref}$  & Output reference \\
$\boldsymbol{y}_c$ & Input to reference model \\
$\boldsymbol{\nu}$  & Pseudo control \\
$\boldsymbol{F_u}$  & Input effectiveness ($\boldsymbol{B}$ matrix in linear system) \\
$\boldsymbol{F_x}$  & System dependence ($\boldsymbol{A}$ matrix in linear system) \\
$\boldsymbol{\Omega}$  & Matrix of actuator bandwidths \\
$\boldsymbol{\Omega_y}$  & Matrix of desired inner-loop bandwidths \\
$\boldsymbol{e_y}$  & Error in the output \\
$\boldsymbol{k_i}$  & Error gain matrix \\
$\boldsymbol{K_i}$  & Error gain matrix in alternative error dynamics \\
$\boldsymbol{I}$   & Identity matrix \\
$\boldsymbol{I}_{m \times m}$   & Identity matrix of size $m$ rows and $m$ columns \\
$\boldsymbol{\Lambda}$   & Input scaling gain matrix \\
$p$  & Roll rate \\
$p_c$  & Roll rate command \\
$L_p$  & Roll damping \\
$L_{p,d}$  & Desired roll damping \\
$L_u$  & Aileron effectiveness \\
$C^{(r)}(D;R)$  & Space of functions differentiable $r$ times with continuous derivatives, $D$ is the domain and $R$ the range.\\
$\boldsymbol{M}^\dagger$  & Any right inverse matrix of $\boldsymbol{M}$ \\
$x^{(r)}$ & derivtive of $x$ with respect to time $r$ times\\
$\cal L$ & Laplace transform\\
$s$ & Laplace variable\\

\multicolumn{2}{@{}l}{Super/Subscripts}\\
$ref$ & From reference model\\
$c$ & Command\\
\multicolumn{2}{@{}l}{Acronyms}\\
NDI & Nonlinear Dynamic Inversion\\
INDI & Incremental Nonlinear Dynamic Inversion\\
ANDI & Actuator-NDI
\end{longtable*}}

\section{Introduction}
\lettrine{O}{ne} of the cornerstones of nonlinear control theory is \ac{NDI}, also called output feedback linearization, which has found applications in many different fields \citep{costa2003,Lane1988,wang2011novel,holzapfel2004dynamic,lombaerts2019nonlinear}.
The concept is based on inverting the nonlinearities of a system, such that the relation between a virtual control input and the output behaves as a linear system, in particular a cascade of integrators.
This transformed system is straightforward to control with a linear control law.
Very complex nonlinear systems can be controlled perfectly through this method in theory, but the results deteriorate if the model is not accurate, if some of the system states cannot be measured accurately or actuator dynamics exist that are not considered \citep{khalil}.

\ac{INDI} is a control method that uses a local linearization of the model to derive a control law to control the defined output and its derivatives, by computation of an increment in the control input, neglecting any state-dependent terms \citep{Bacon2000}.
Through the feedback of derivatives of the system output, such as angular acceleration in the case of inner loop control of an aircraft, unmodeled effects, and disturbances are directly measured, and compensated for in the next control increment, which led to an increased popularity of the concept in flight control applications \citep{steffensen2022longitudinal,Myschik, UniStuttgart,GABRYS2019429,huang2018robust,grondman2018design,li2021extended}.
In \citep{smeur2016} it is shown that under certain simplifying assumptions, the closed-loop system responds to disturbances or unmodeled dynamics with the combined dynamics of the actuators and any filtering that is done on the output.
This has also been observed in practical experiments as well \citep{smeur2016}.

It would seem that if one were to include the partial derivatives of the states with respect to the states in the INDI control law,  one would obtain better performance in both tracking and disturbance rejection.
\citet{Wang2019} suggest keeping a term with the state change "$\Delta \boldsymbol{x}$" over one time step as part of the control law.
The drawback of this approach is that when actuator dynamics cannot be neglected, the computed increment in input is not realized within one time sample.
That makes the considered anticipation of the state change insufficient.
In fact, this problem is hard to solve if one follows the 'traditional' derivation of the \ac{INDI} control law, as will be further detailed in Section \ref{sec:indi}.
\citet{Li2018} propose to add a large gain, such that if the actuator behaves as a first-order system, the system will achieve the desired value within one controller time step, effectively removing any actuator dynamics.
This approach is not realistic, as controllers typically run at a frequency much higher than the bandwidth of the actuators.

\citet{zhou2021} describe a method of including state dependent terms in the INDI control law in discrete time.
The drawback of this approach is that, due to the discrete formulation, a control input is calculated that will solve for the virtual control within one time step.
In most cases, the time constant of the actuators is larger than one time step of the controller, which would lead to large inputs to the actuators and possibly hidden oscillations between samples.

Several concepts have been proposed, that scale the control effectiveness matrix, for different reasons.
\citet{cordeiro2019cascaded} noted that an input gain scaling, which was used to reduce the influence of noise, can reduce the closed loop bandwidth.
In \citep{cordeiro2021increased}, it was shown that the input scaling also increases robustness with respect to time delays.
\citet{pfeifle2022time} proposed an additional gain that depends on the sampling time and the actuator time constant based on an alternative derivation of the incremental dynamic inversion control law.
\citet{Raab2019} related the meaning and value of an input gain scaling to the actuator parameters.

\citet{Raab2019} also suggested that actuator dynamics can be included in the derivation of the controller, by taking an additional derivative of the system output.
Essentially, this is achieved by considering the actuator dynamics as a part of the system dynamics, hence the control law then also inverts the actuator dynamics.
The main benefits of this approach are the ability to incorporate actuator rate constraints in the control allocation and artificially choose faster or slower effective actuator dynamics, which is especially useful if actuators with different effective bandwidth are used to control coupled outputs.
However, the state dependent effects are not taken into account in \citet{Raab2019}.

In an extension, \citet{Bhardwaj2021} based a reference model design on dynamic inversion including the actuator dynamics.
From this physical reference model a feed-forward control signal is generated, that accounts for state-dependent model effects.
This theoretically leads to perfect tracking of the reference model, if the reference model equals the plant and if the states of the reference model equal the plant states, e.g. if there are no model uncertainties and no disturbances.
However, the benefit of this approach is deteriorated if the plant is not exactly following the reference model.
If the reference model states are different from the plant states, the feed-forward command will not be correct, and this will lead to unpredictable error dynamics.
Furthermore, the plant states that are not controlled (zero dynamics), might differ from the corresponding states of the reference model, in case that they are directly influenced by certain control effectors. In case this influence is incorporated in the reference model some sort of feedback from the Control Allocation is required to make sure that these uncontrolled states do not diverge from the respective plant states, such that an accurate feed forward command can be achieved, which has not been shown how to be accomplished. 

The contribution of this paper is based on an alternative derivation of a set of incremental control laws using nonlinear dynamic inversion.
In particular:
\begin{enumerate}
    \item Formulation of a novel sensor based incremental nonlinear dynamic inversion control law taking into account state dependent terms in a consistent and straightforward manner.
    \item Formulation of error dynamics split into actuation and system dynamics. This allows specifying actuation dynamics according to actuator limitations and system dynamics according to vehicle eigenresponse requirements.
    \item An alternative derivation of the incremental nonlinear dynamic inversion control law is presented based on the above mentioned error dynamics. The derivation shows that in case the actuators can be assumed to be fast in relation to the state dynamics the traditional INDI law approximates the derived exact NDI control law.
    \item In addition, an alternative derivation of incremental nonlinear dynamic inversion with input scaling gain is presented together with an associated interpretation of inner-loop bandwidth reduction/increase.
\end{enumerate} 

In Section \ref{sec:ndi} we derive the proposed incremental \ac{NDI} control law for a nonlinear system with first-order linear actuators. In Section \ref{sec_erdyn} we derive the resulting control law with error dynamics split between actuation and system dynamics. In Section \ref{sec:indi} we compare the conventional INDI and INDI with input scaling gain to the derived NDI control law. In Section \ref{sec_INDI_w_actuators}, we compare the derived INDI control law to existing INDI control laws that take into account the actuator dynamics. In Section \ref{sec_example} we compare the derived NDI with the INDI control laws on a simple example of the rolling motion of a fixed wing aircraft. In Section \ref{sec_discussion} we discuss the results and in Section \ref{sec_conclusion} we give the concluding remarks.


\section{Nonlinear Dynamic Inversion control law with first-order linear actuator dynamics} \label{sec:ndi} 
To introduce the concept, consider the general system with first-order linear actuator dynamics given by:
\begin{equation}
\label{eq_siso_system_dynamics}
\begin{array}{l}
\boldsymbol{\dot {x}} = \boldsymbol{f}(\boldsymbol{x},\boldsymbol{u})\\
\boldsymbol{y} = \boldsymbol{h}(\boldsymbol{x}),
\end{array}
\end{equation}
where $\boldsymbol{x} \in \mathbb{R}^n$ is the system state, $\boldsymbol{u} \in \mathbb{R}^k$ is the actuator state, $\boldsymbol{y} \in \mathbb{R}^m$ is the output, $\boldsymbol{f} \in C^{(r+1)}(\mathbb{R}^n \times \mathbb{R}^k;\mathbb{R}^n)$ and $\boldsymbol{h} \in C^{(r+1)}(\mathbb{R}^n;\mathbb{R}^m)$. The actuator dynamics are given by:
\begin{equation}
    \boldsymbol{\dot u} = \boldsymbol{\Omega} \left( \boldsymbol{u_c} - \boldsymbol{u} \right)
    \label{eq_1storder_act}
\end{equation}
where $\boldsymbol{\Omega} \in \mathbb{R}^{(k \times k)}$ is a diagonal matrix with constant elements representing the bandwidth of the different actuators. Assume the system to have a relative degree of $r \in \mathbb{N}$ with respect to $\boldsymbol{u}$ for all elements of $\boldsymbol{y}$, i.e. $r$ is the number of times the output $\boldsymbol{y}$ has to be differentiated with respect to time for the input $\boldsymbol{u}$ to appear explicitly. The $r$'th derivative can be expressed as follows:
\begin{equation}
\label{eq_y_r}
{\boldsymbol{y}^{(r)}} = \boldsymbol{F}(\boldsymbol{x},\boldsymbol{u})
\end{equation}
with $\boldsymbol{F} \in C^{1}(\mathbb{R}^n \times \mathbb{R}^k;\mathbb{R}^m)$. 
Traditionally, this expression for the $r$'th derivative of $\boldsymbol{y}$ is used to obtain the control law by inverting for $\boldsymbol{u}$ and by relating $\boldsymbol{y}^{(r)}$ to the virtual control $\boldsymbol{\nu}$, see for example \cite{khalil}.
In those cases, actuator dynamics are not considered, or assumed to be relatively fast such that they can be neglected. Instead, in this paper the nonlinear dynamic inversion is continued through the actuator dynamics by performing one more differentiation w.r.t. time:
\begin{equation}
{\boldsymbol{y}^{(r + 1)}} = {\boldsymbol{F_x}} \boldsymbol{\dot x} + {\boldsymbol{F_u}} \boldsymbol{\dot u}
\label{eq_siso_ynp1}
\end{equation}
where ${\boldsymbol{F}_x} \coloneqq   \frac{{\partial \boldsymbol{F}(\boldsymbol{x},\boldsymbol{u})}}{{\partial \boldsymbol{x}}}$ and ${\boldsymbol{F}_u} \coloneqq   \frac{{\partial \boldsymbol{F}(\boldsymbol{x},\boldsymbol{u})}}{{\partial \boldsymbol{u}}}$. Now the actuator relation in Eq. \eqref{eq_1storder_act} is substituted into Eq. \eqref{eq_siso_ynp1} to obtain a relation for $\boldsymbol{y}^{(r+1)}$ that includes the actuator command.
\begin{equation}
    {\boldsymbol{y}^{(r + 1)}} = {\boldsymbol{F_x}} \boldsymbol{\dot x} + {\boldsymbol{F_u}} \boldsymbol{\Omega} \left( \boldsymbol{u_c} - \boldsymbol{u} \right)
    \label{eq_siso_ypp1}
\end{equation}
Note, that traditionally the relation for $\boldsymbol{\dot{x}}$, would be substituted in into Eq. \eqref{eq_siso_ypp1}. Here, we keep $\boldsymbol{\dot{x}}$ to allow it to be possibly determined by measurements. Choose now  ${\boldsymbol{y}^{(r + 1)}}=\boldsymbol{\nu}$, where $\boldsymbol{\nu}$ is a virtual control command. If it is assumed that ${\boldsymbol{F}_u} \boldsymbol{\Omega}$ has full row rank, the following choice of $\boldsymbol{u_c}$ will make ${\boldsymbol{y}^{(r + 1)}}=\boldsymbol{\nu}$:
\begin{equation}
\boldsymbol{u_c} = (\boldsymbol{F_u}\boldsymbol{\Omega} )^{\dagger}\left(  \boldsymbol{\nu} - \boldsymbol{F_x} \boldsymbol{\dot x} \right) + \boldsymbol{u}
\label{eq_siso_ctrl_law}
\end{equation}
where $({\boldsymbol{F}_u}\Omega )^{\dagger}$ denotes a right inverse matrix that solves the linear equation system given in Eq. \eqref{eq_siso_ypp1}.
The control signal $\boldsymbol{u_c}$ linearizes the response from the virtual control input $\boldsymbol{\nu}$ to the output $\boldsymbol{y}$ to a chain of integrators.

Often a linear controller is used to regulate the output $\boldsymbol{y}$. In this paper, for the next step, the virtual control input $\boldsymbol{\nu}$ is designed using a linear error controller such that the specified error dynamics are achieved. 
The order of the error dynamics in $\boldsymbol{y}$ will correspond to the sum of the relative degree and the order of the actuator dynamics, such that the error dynamics could be specified by
\begin{equation}
\boldsymbol{e_y}^{(r+1)}+\sum\limits_{i =0}^{ r } {{\boldsymbol{k}_i}\boldsymbol{e_y}^{(i)}}=0
\label{eq_error_dyn_simple}
\end{equation}
where $\boldsymbol{e_y} = \boldsymbol{y}_{ref} - \boldsymbol{y}$ is the error in $\boldsymbol{y}$, $\boldsymbol{e_y}^{(i)}$ is the $i$'th derivative with respect to time, and $\boldsymbol{y}^{(r+1)}$ is chosen as the virtual control $\boldsymbol{\nu}$ such that:
\begin{equation}
    \boldsymbol{\nu}  
    = \boldsymbol{y}_{ref}^{(r + 1)} + \sum\limits_{i = 0}^r {{\boldsymbol{k}_i}\boldsymbol{e_y}^{(i)}} 
\end{equation}
Inserting $\boldsymbol{\nu}$ into Eq. \eqref{eq_siso_ctrl_law} results in the control law:
\begin{equation}
\boldsymbol{u_c} = (\boldsymbol{F_u}\boldsymbol{\Omega} )^{\dagger}\left( { \boldsymbol{y}_{ref}^{(r+1)}+\sum\limits_{i =0}^{ r } {{\boldsymbol{k}_i}\boldsymbol{e_y}^{(i)}}- \boldsymbol{F_x}\boldsymbol{\dot x}} \right) + \boldsymbol{u}
\label{eq_final_ctrl_NDI_law}
\end{equation}
The resulting control law is a NDI control law with linear error controller, where the actuator dynamics were additionally included in the system dynamics and inverted.
By inserting Eq. \eqref{eq_final_ctrl_NDI_law} into Eq. \eqref{eq_siso_ypp1}, the desired error dynamics are obtained as designed, because $\boldsymbol{F_u}\boldsymbol{\Omega}$ has full row rank such that $(\boldsymbol{F_u}\boldsymbol{\Omega})(\boldsymbol{F_u}\boldsymbol{\Omega})^{\dagger}=\boldsymbol{I}_{m \times m}$:
\begin{equation}
\begin{split}
    {\boldsymbol{y}^{(r + 1)}} = \boldsymbol{F_x}\boldsymbol{\dot x} + \boldsymbol{F_u}\boldsymbol{\Omega}(\boldsymbol{F_u}\boldsymbol{\Omega} )^\dag   {\left(  { \boldsymbol{y}_{ref}^{(r+1)}+\sum\limits_{i =0}^{ r } {{\boldsymbol{k}_i}\boldsymbol{e_y}^{(i)}}- \boldsymbol{F_x}\boldsymbol{\dot x}}  \right)}
    =  { \boldsymbol{y}_{ref}^{(r+1)}+\sum\limits_{i =0}^{ r } {{\boldsymbol{k}_i}\boldsymbol{e_y}^{(i)}}}  \\
\end{split}
\label{eq_error_dynamics_yrplusone}
\end{equation}
This leads to tracking of the corresponding derivative of a reference model ${\boldsymbol{y}_{ref}}^{(r+1)}$, and the resulting error dynamics will correspond to Eq. \eqref{eq_error_dyn_simple}.

This control law can be compared to the control lawproposed by \citet{Bhardwaj2021}, where the state dependent term $\boldsymbol{F_x}\boldsymbol{\dot{x}}$ in the control law in Eq. \eqref{eq_final_ctrl_NDI_law}, is generated by a physical reference model using an additional feed forward term corresponding to $\boldsymbol{F_x} \boldsymbol{\dot {x}}_{ref}$.
In that case, the feed forward from the reference model is only correct if the system is tracking the reference model and the states corresponding to the zero dynamics are matching the respective plant states.
In addition, in case of disturbances which lead to a perturbation with differences between the desired and actual trajectory, the actual error dynamics will differ from the desired error dynamics, because they will be excited by the term $\boldsymbol{F_x}(\boldsymbol{\dot{x}}-\boldsymbol{\dot{x}}_{ref})$ as can be seen in Eq. \eqref{eq_error_dynamics_yrplusone}. \footnote{Another subtle issue that can arise is if the system $B$ matrix and the $F_u$ matrix used for control allocation are different, because the system contains states that are not directly controlled by the inversion law, then the dynamics of the states in the reference model that are not controlled can depend on the control allocation.}
However, in the proposed inversion based control law, where the state derivatives are used, the error dynamics will correspond to the desired error dynamics.
A possible drawback of this approach could be reduced stability margins when analyzing the linearized closed loop system due to the additional feedback, but this requires further investigation.
It will become apparent in section \ref{sec:indi} that the classical \ac{INDI} approximates this \ac{NDI} law under the condition that the actuators are fast with equal bandwidth.

\subsection{Choice of Desired Error Dynamics}\label{sec_erdyn}
In this section we propose a particular structure of the error dynamics in order to split up the error tracking in terms of actuators and system dynamics. This new formulation also serves effective in deducing the approximations required to obtain the popular INDI control law, as will be shown in section \ref{sec:indi}. 
The error dynamics in Eq. \eqref{eq_error_dyn_simple} have a generic form, and is shaped by choice of the gains $k_i$. Some choices of these gains could lead to error dynamics that cannot be realistically obtained with real world actuators, due to physical constraints in the actuators.
It therefore makes sense to design the error controller such as to include actuator dynamics with a bandwidth equal or comparable to the real actuators. Furthermore, as will be shown in section \ref{sec:indi}, the proposed formulation of the error dynamics allows a direct comparison to the INDI control law.

The order of the error dynamics in $\boldsymbol{y}$ corresponds to the sum of the relative degree of $\boldsymbol{y}$ and the order of the actuator dynamics.
The desired dynamics can therefore be interpreted as cascaded dynamics composed of the slower desired system error dynamics and a faster desired inner loop dynamics with bandwidth $\boldsymbol{\Omega_y}$, corresponding to the actuator dynamics.
It can hence make sense to express the desired error dynamics Eq. \eqref{eq_error_dyn_simple} in the Laplace domain as the product of dynamics due to the system physics and due to the actuators as follows:
\begin{equation}
{\boldsymbol{E_y}(s)}\left( {{s^r} \boldsymbol{I} + \sum\limits_{i = 0}^{r - 1} {{\boldsymbol{K}_i}{s^i}} } \right)\left( s\boldsymbol{I} + \boldsymbol{\Omega_y} \right) = 0
\label{eq_combined_e_dyn}
\end{equation}
where $\boldsymbol{e_y} = \boldsymbol{y}_{ref} - \boldsymbol{y}$ is the error in $\boldsymbol{y}$ with Laplace transform $\boldsymbol{E_y}(s)$, $\left( {{s^r} \boldsymbol{I} + \sum\limits_{i = 0}^{r - 1} {{\boldsymbol{K}_i}{s^i}} } \right)$ describes the desired error dynamics with respect to the system, $\left( {s\boldsymbol{I} + {\boldsymbol{\Omega _y}}} \right)$ is the desired error dynamics due to the first-order actuators, and $\boldsymbol{K}_i$ are the new error controller gains defining the desired system error dynamics. The system error dynamics $\boldsymbol{E_y}(s)\left( {{s^r} \boldsymbol{I} + \sum\limits_{i = 0}^{r - 1} {{\boldsymbol{K}_i}{s^i}} } \right)$ can be formulated in the time domain as 

\begin{equation}
{{\cal L}^{ - 1}}\left( {\boldsymbol{E_y}(s)\left( {{s^r} \boldsymbol{I} + \sum\limits_{i = 0}^{r - 1} {{\boldsymbol{K}_i}{s^i}} } \right)} \right) = \boldsymbol{e_y}^{(r)} + \sum\limits_{i = 0}^{r-1} {{\boldsymbol{K}_i}\boldsymbol{e_y}^{(i)}}  = 0
    \label{eq_system_ed}
\end{equation}
where $\boldsymbol{K}_i$ are the error gain matrices and $\mathcal{L}^{ - 1}$ is the inverse Laplace transform. Using Eq. \eqref{eq_system_ed}, the combined error dynamics from Eq. \eqref{eq_combined_e_dyn} can be described in the time domain by:
\begin{equation}
\begin{split}
 \frac{d}{{dt}}\left( {\boldsymbol{e_y}^{(r)} + \sum\limits_{i = 0}^{r-1} {\boldsymbol{K}_{i}\boldsymbol{e_y}^{(i)}} } \right) + \boldsymbol{\Omega _y}\left( {\boldsymbol{e_y}^{(r)} + \sum\limits_{i = 0}^{r-1} {\boldsymbol{K_i}\boldsymbol{e_y}^{(i)}} } \right) 
= {\boldsymbol{e_y}^{(r+1)} + \sum\limits_{i = 1}^{r} {{\boldsymbol{K}_{i-1}}\boldsymbol{e_y}^{(i)}} }  + \boldsymbol{\Omega _y}\left( {\boldsymbol{e_y}^{(r)} + \sum\limits_{i = 0}^{r-1} {{\boldsymbol{K}_i}\boldsymbol{e_y}^{(i)}} } \right) =\boldsymbol{0}
\end{split}
\label{eq_td_dyn}
\end{equation}
Note that if a formulation as given in Eq. \eqref{eq_error_dyn_simple} is preferred, Eq. \eqref{eq_td_dyn} can be reformulated as
\begin{equation}
\begin{split}
     \boldsymbol{e_y}^{(r+1)} +(\boldsymbol{K}_{r-1}+\boldsymbol{\Omega_y})\boldsymbol{e_y}^{(r)}  +\sum\limits_{i = 1}^{r-1} \left( ({\boldsymbol{K}_{i-1}+\boldsymbol{\Omega_y} \boldsymbol{K}_i})\boldsymbol{e_y}^{(i)}\right)  + {\boldsymbol{\Omega _y} \boldsymbol{K}_0}\boldsymbol{e_y}=\boldsymbol{0}
\label{eq_errordyn_exact}
\end{split}
\end{equation}
and the gain matrices $\boldsymbol{k}_i$  can be obtained directly.
Inserting $\boldsymbol{e_{y}}^{(r+1)} = \boldsymbol{y}_{ref}^{(r+1)} - \boldsymbol{y}^{(r+1)}$ into Eq. \eqref{eq_td_dyn} and solving for ${\boldsymbol{y}^{(r + 1)}}$ provides the pseudo control input $\boldsymbol{\nu}$ that can be used in Eq. \eqref{eq_siso_ctrl_law}:
\begin{equation}
\begin{split}
    \boldsymbol{\nu} \mathop = {\boldsymbol{y}_{ref}}^{(r + 1)} + \sum\limits_{i = 1}^r {{\boldsymbol{K}_{i-1}}\boldsymbol{e_y}^{(i)}}  + \boldsymbol{\Omega _y}\left( {\boldsymbol{e_y}^{(r)} + \sum\limits_{i = 0}^{r-1} {{\boldsymbol{K}_i}\boldsymbol{e_y}^{(i)}} } \right)
    \end{split}
\label{eq_error_dyn_siso}
\end{equation}
with reference dynamics given in terms of the respective derivatives of $y_{ref}$, and error controller gains given by $\Omega _y$ and $K_i$. The final control law is then given by
\begin{equation}
\begin{split}
 \boldsymbol{u_c} = {(\boldsymbol{F_u}\boldsymbol{\Omega} )^\dag } \Biggl( {\boldsymbol{y}_{ref}}^{(r + 1)} + \sum\limits_{i = 1}^r {{\boldsymbol{K}_{i-1}}\boldsymbol{e_y}^{(i)}} 
 +\boldsymbol{\Omega _y}\left( {\boldsymbol{e_y}^{(r)} + \sum\limits_{i = 0}^{r-1} {{\boldsymbol{K}_i}\boldsymbol{e_y}^{(i)}} } \right) - \boldsymbol{F_x}\boldsymbol{\dot x} \Biggr) + \boldsymbol{u}
\end{split}
\label{eq_siso_ctrl_law_final_complex}
\end{equation}

The resulting error dynamics will correspond to Eq. \eqref{eq_combined_e_dyn}, which can be interpreted as cascaded dynamics composed of the slower desired system dynamics and a faster desired inner loop dynamics bandwidth $\boldsymbol{\Omega_y}$, due to the actuator dynamics.
Since Eq. \eqref{eq_siso_ctrl_law_final_complex} essentially inverts the actuator dynamics as well, the system could be made arbitrarily fast through the choice of $\boldsymbol{\Omega_y}$.
However for practical applications, the actuators were designed to operate up to a certain bandwidth, and the choice $\boldsymbol{\Omega_y}$ should be limited to the design bandwidth of the actuators accordingly. In some cases, the dynamics of a control effector is limited by the effector dynamics and not the physical actuator. Here an increase in effector bandwidth can be obtained by selecting $\boldsymbol{\Omega_y}$ larger than the actuator bandwidth in the considered direction.
The second degree of freedom consists of the choice of the error gain matrices $\boldsymbol{K}_i$'s. The $\boldsymbol{K}_i$ matrices should be chosen to satisfy requirements on the error dynamics related to the system dynamics. This will be demonstrated in the example in Section \ref{sec_example}, where the roll damping is increased for a fixed wing aircraft. At last, the closed-loop dynamics from reference input $\boldsymbol{y}_c$ to response $\boldsymbol{y}$ is determined by the choice of reference model dynamics, i.e. the dynamics from $\boldsymbol{y}_c$ to $\boldsymbol{y}_{ref}$. 

\section{Comparison with Incremental Nonlinear Dynamic Inversion} \label{sec:indi}
This section compares the \ac{NDI} control law, that was derived in section \ref{sec:ndi}, to an \ac{INDI} controller, which has been derived for example by \citet{Bacon2000} and \citet{sieberling2010robust}.
The INDI controller will be derived along the same lines, keeping the nomenclature the same as in the previous section, such that the controllers can be effectively compared. The classic INDI derivation contains some inaccuracies, which are pointed out in this section as well.


Again, consider the system in Eq. \eqref{eq_siso_system_dynamics}, and the $r$'th derivative of the output as in Eq. \eqref{eq_y_r}.
Take the Taylor expansion of $\boldsymbol{F}(\boldsymbol{x},\boldsymbol{u})$ with respect to $\boldsymbol{x}$ and $\boldsymbol{u}$:
\begin{equation}
\begin{split}
    {\boldsymbol{y}^{(r)}}(t) =  \boldsymbol{F}(\boldsymbol{x}(t),\boldsymbol{u}(t))  =  \boldsymbol{F}({\boldsymbol{x}_0},{\boldsymbol{u}_0}) + \boldsymbol{F_x}\left( {\boldsymbol{x}(t) - {\boldsymbol{x}_0}} \right) + \boldsymbol{F_u}(\boldsymbol{u}(t) - {\boldsymbol{u}_0})  + O\left( {\Delta \boldsymbol{x},\Delta \boldsymbol{u}} \right)
\end{split}
\label{eq:taylor_exp}
\end{equation}
where $\boldsymbol{F_x} \coloneqq   \frac{{\partial \boldsymbol{F}(\boldsymbol{x},\boldsymbol{u})}}{{\partial \boldsymbol{x}}}$ and $\boldsymbol{F_u} \coloneqq   \frac{{\partial \boldsymbol{F}(\boldsymbol{x},\boldsymbol{u})}}{{\partial \boldsymbol{u}}}$ and $O$ denotes higher order terms.
The term $\boldsymbol{F}\left( {{\boldsymbol{x}_0},{\boldsymbol{u}_0}} \right)$ can also be denoted by ${\boldsymbol{y}_0}^{(r)}$.
It should be noted that $\boldsymbol{x}$ and $\boldsymbol{u}$ are functions of time. The Taylor expansion in Eq. \eqref{eq:taylor_exp} is performed with respect to $\boldsymbol{x}$ and $\boldsymbol{u}$. For $\boldsymbol{x}_0$ and $\boldsymbol{u}_0$ it makes sense to choose a state and control input, corresponding to the system, a small time instance ago, i.e. define:
\begin{equation}
\label{eq:dt}
    \begin{split}
{\boldsymbol{x}_0} & = \boldsymbol{x}(t - \Delta t)\\
{\boldsymbol{u}_0} & = \boldsymbol{u}(t - \Delta t)\\
\end{split}
\end{equation}
If ${\boldsymbol{y}^{(r)}}$ is chosen as virtual control $\boldsymbol{\nu}$ and $\boldsymbol{u}$ is chosen as the control signal $\boldsymbol{u_c}$, the following relation is obtained:
\begin{equation}
\boldsymbol{\nu}  = {\boldsymbol{y}_0}^{(r)} + \boldsymbol{F_x}\left( {\boldsymbol{x} - {\boldsymbol{x}_0}} \right) + \boldsymbol{F_u}\left( {\boldsymbol{u_c} - {\boldsymbol{u}_0}} \right) + O(\Delta \boldsymbol{{x}},\Delta \boldsymbol{{u}})
\label{eq_nu_relation_siso_INDI}
\end{equation}
It is assumed that the higher order terms can be neglected.
The control law is obtained by solving Eq. \eqref{eq_nu_relation_siso_INDI} for $\boldsymbol{u_c}$:
\begin{equation}
\boldsymbol{u_c} = \boldsymbol{F_u}^{\dagger}\left( {\boldsymbol{\nu}  - {\boldsymbol{y}^{(r)}_0} - \boldsymbol{F_x}\left( {\boldsymbol{x} - {\boldsymbol{x}_0}} \right)} \right) + \boldsymbol{u_0}
\label{eq_extact_INDI_control_law}
\end{equation}
The term $\boldsymbol{F_x}(\boldsymbol{x}-\boldsymbol{x}_0)$ is usually neglected, with several different arguments involving the bandwidth of the actuator and a sufficiently small sample time \cite{sieberling2010robust,Tal2019}, leading to the following control law:
\begin{equation}
\boldsymbol{u_c} = \boldsymbol{F_u}^{\dagger}\left( {\boldsymbol{\nu}  - {\boldsymbol{y}^{(r)}_0} } \right) + \boldsymbol{u_0}
\label{eq_INDI_control_law_classic}
\end{equation}
If the term $\boldsymbol{F_x}(\boldsymbol{x}-\boldsymbol{x}_0)$, or some approximation of this is sought to be included, it is not clear how $\Delta \boldsymbol{x} $ in Eq. \eqref{eq_extact_INDI_control_law} should be chosen. \citet{Wang2019} proposed to use $\Delta \boldsymbol{x} = \boldsymbol{\dot{x}}(t-\Delta t) \Delta t$ with $\Delta t$ chosen as the sample time. 
However, because the actuator command, Eq. \eqref{eq_extact_INDI_control_law}, will not be instantaneously reached by the actuator, due to the actuator dynamics, the $\boldsymbol{F}_x(\boldsymbol{x}-\boldsymbol{x}_0)$ term in the $\boldsymbol{y}^{(r)}$ dynamics of Eq. \eqref{eq:taylor_exp} will not be exactly canceled by the control signal $u(t)$.
The resulting stability properties of the INDI control law are unclear, neither is it clear what was neglected from the Taylor expansion.
The virtual control $\boldsymbol{\nu}$ in Eq. \eqref{eq_INDI_control_law_classic}, is derived by specifying the desired error dynamics in $\boldsymbol{y}$. Usually, the desired error dynamics are specified by Eq. \eqref{eq_system_ed}, such that the virtual control is given by
\begin{equation}
\boldsymbol{\nu}  =  {{\boldsymbol{y}_{ref}}^{(r)} + \sum\limits_{i = 0}^{r-1} {{\boldsymbol{K}_i} {\boldsymbol{e_y}^{(i)}} } }
\label{eq:virtual_ctrl_indi}
\end{equation}
Combining Eq. \eqref{eq_INDI_control_law_classic} and Eq. \eqref{eq:virtual_ctrl_indi}, the following INDI control law is found:
\begin{equation}
\begin{split}
    \boldsymbol{u_c} = \boldsymbol{F_u}^{\dagger}\left( {\boldsymbol{y}_{ref}}^{(r)} - {\boldsymbol{y}^{(r)}_0} +  \sum\limits_{i = 0}^{r-1} {{\boldsymbol{K}_i} {\boldsymbol{e_y}^{(i)}} }  \right) + {\boldsymbol{u}_0} 
    = \boldsymbol{F_u}^{\dagger}\left( \boldsymbol{e_y}^{(r)} +  \sum\limits_{i = 0}^{r-1} {{\boldsymbol{K}_i} {\boldsymbol{e_y}^{(i)}} } \right) + {\boldsymbol{u}_0}
\end{split}
\label{eq:indi_with_dx}
\end{equation}

By comparing the above control law with the exact \ac{NDI} law in Eq. \eqref{eq_siso_ctrl_law_final_complex}, it is possible to identify which part was neglected.
In the following, it is shown that under certain conditions relating to the actuators and the NDI control law parameters, the INDI law approximates the exact NDI law.
Assume that:
\begin{enumerate}
    \item All actuators have the same bandwidth, i.e. $\boldsymbol{\Omega}=\omega \boldsymbol{I}_{k\times k}$, where $\omega$ is a positive scalar.
    \item Choose $\boldsymbol{\Omega_y}=\omega \boldsymbol{I}_{m\times m}$.
\end{enumerate}
Then
\begin{equation}
\begin{split}
        {(\boldsymbol{F_u}\boldsymbol{\Omega} )^\dag } \boldsymbol{\Omega_y} = {(\boldsymbol{F_u}\omega {\boldsymbol{I}_{k \times k}})^\dag }\omega {\boldsymbol{I}_{m \times m}} 
        = {(\boldsymbol{F_u}\omega )^\dag }\omega  ={(\boldsymbol{F_u} )^\dag }\frac{1}{\omega}\omega ={(\boldsymbol{F_u})^\dag }
\end{split}
\end{equation}
Hence, the control law from Eq. \eqref{eq_siso_ctrl_law_final_complex} evaluates to:
\begin{equation}
\begin{split}
    \boldsymbol{u_c} = \frac{1}{\omega}{(\boldsymbol{F_u})^{ \dag}}\left( {\boldsymbol{y}_{ref}^{(r + 1)} + \sum\limits_{i = 1}^r {{\boldsymbol{K}_{i-1}}} \boldsymbol{e_y}^{(i)} - \boldsymbol{F_x}\boldsymbol{\dot x}{\rm{ }}} \right) 
+ {(\boldsymbol{F_u})^{ \dag}}\left( {\boldsymbol{e_y}^{(r)} + \sum\limits_{i = 0}^{r-1} {{\boldsymbol{K}_i}\boldsymbol{e_y}^{(i)}} } \right){\rm{ }} + \boldsymbol{u}
\end{split}
\label{eq_siso_ctrl_law2}
\end{equation}
It is seen here that in the limit where the actuator bandwidth $\omega$ approaches infinity, the \ac{NDI} law turns exactly into the \ac{INDI} control law of Eq. \eqref{eq:indi_with_dx}:
\begin{equation}
\begin{split}
    \boldsymbol{u_c}  =& \mathop {\lim }\limits_{\omega  \to \infty } \Biggl( \frac{1}{\omega}{(\boldsymbol{F_u})^{ \dag}}\left( {\boldsymbol{y}_{ref}^{(r + 1)} + \sum\limits_{i = 1}^r {{\boldsymbol{K}_{i-1}}} \boldsymbol{e_y}^{(i)} - \boldsymbol{F_x}\boldsymbol{\dot x}{\rm{ }}} \right) 
     + {(\boldsymbol{F_u})^{ \dag}}\left( {\boldsymbol{e_y}^{(r)} + \sum\limits_{i = 0}^{r-1} {{\boldsymbol{K}_i}\boldsymbol{e_y}^{(i)}} } \right){\rm{ }} + \boldsymbol{u} \Biggr) \\
 =& (\boldsymbol{F_u})^{\dagger}\left( \boldsymbol{e_y}^{(r)} +  \sum\limits_{i = 0}^{r-1} {{\boldsymbol{K}_i} {\boldsymbol{e_y}^{(i)}} } \right) + \boldsymbol{u}
\end{split}
\label{eq:comparison}
\end{equation}
The \ac{NDI} law of Section \ref{sec:ndi} perfectly inverts the system, so by comparing the INDI law given in Eq. \eqref{eq:indi_with_dx} with the exact sensor based NDI law given by Eq. \eqref{eq_siso_ctrl_law2} reveals exactly which part is neglected by applying the INDI control law, namely the first term in Eq. \eqref{eq_siso_ctrl_law2}.
These missing terms will lead to errors in reference model tracking, and lead to error dynamics that are different from the designed error dynamics. It is seen that the mismatch between the INDI law and the exact sensor based NDI law is vanishing for sufficiently large actuator bandwidth $\omega$.\\
\\
This comparison shows that the derivation that was performed in Section \ref{sec:ndi} is a useful alternative to arrive at the incremental nonlinear dynamic inversion law as it does not require any ad-hoc arguments.
As such, the derivation provides new theoretical support for the \ac{INDI} control method, while also providing a means to compensate for model-dependent terms ($\boldsymbol{F_x} \boldsymbol{\dot{x}}$) in the control law.
In INDI, these terms are not compensated for, and depending on the system, this may lead to significant errors in tracking and unpredictable error dynamics.

As shown in the Appendix, the \ac{INDI} law does not hold in the limit, i.e. it is only valid for $\omega<\infty$, hence the above equation essentially states that for a fixed time, the \ac{INDI} law approximates the true \ac{NDI} law arbitrarily well by the choice of sufficiently high actuator bandwidth.

It can further be shown that the input scaling as suggested by \citet{cordeiro2019cascaded} and \citep{cordeiro2021increased} for reducing the influence of noise and increasing the robustness with respect to time delays, can be interpreted as a modification of the innermost bandwidth of $\boldsymbol{y}$: $\boldsymbol{\Omega_y}=\boldsymbol{\Lambda} \omega$, where the input scaling gain matrix $\boldsymbol{\Lambda} \in \mathbb{R}^{m\times m}$ is diagonal.
If this relation is inserted into the NDI control law in Eq. \eqref{eq_siso_ctrl_law_final_complex}, and it is assumed that the actuators have equal bandwidth, i.e. $\boldsymbol{\Omega}=\omega \boldsymbol{I}_{k\times k}$, and taking the limit of the bandwidth $\omega$ going to infinity, we obtain the conventional INDI control law with the scaling gain matrix that was proposed by \citet{cordeiro2019cascaded}:
\begin{equation}
\begin{split}
    \boldsymbol{u_c}  =& \mathop {\lim }\limits_{\omega  \to \infty } \Biggl( \frac{1}{\omega }{{(\boldsymbol{F_u})}^{{\rm{ \dagger}}}}\left( {\boldsymbol{y}_{ref}^{(r + 1)} + \sum\limits_{i = 1}^r {{\boldsymbol{K}_{i-1}}} \boldsymbol{e_y}^{(i)} - \boldsymbol{F_x}\boldsymbol{\dot x}} \right) 
    + \frac{1}{\omega }{{(\boldsymbol{F_u})}^{{\rm{ \dagger}}}}\boldsymbol{\Lambda} \omega \left( {\boldsymbol{e_y}^{(r)} + \sum\limits_{i = 0}^{r - 1} {{\boldsymbol{K}_i}\boldsymbol{e_y}^{(i)}} } \right) + \boldsymbol{u} \Biggr) \\
 = & (\boldsymbol{F_u})^{\dagger}\boldsymbol{\Lambda}\left( \boldsymbol{e_y}^{(r)} +  \sum\limits_{i = 0}^{r-1} {{\boldsymbol{K}_i} {\boldsymbol{e_y}^{(i)}} } \right) + \boldsymbol{u}
\end{split}
\end{equation}
\citet{cordeiro2019cascaded} observed that choosing the diagonal elements of $\boldsymbol{\Lambda}$ smaller than 1 reduces the  closed-loop bandwidth.
With this formulation of the control law, the scaling gain can be directly identified as a modification of the desired innermost bandwidth.
If the scaling factor is 1, as in conventional INDI, this corresponds to a desired innermost bandwidth equal to the actuator bandwidth.

\section{Comparison with Incremental Nonlinear Dynamic Inversion including Actuators} \label{sec_INDI_w_actuators}
\citet{smeur2016} showed that when actuator dynamics are present and the INDI control law in Eq. \eqref{eq_INDI_control_law_classic} is applied, then
$\boldsymbol{y}^{(r)}$ follows $\boldsymbol{\nu}$ with the dynamics of the actuator, in the case that the term $\boldsymbol{F_x} \boldsymbol{\dot x}$ in Eq. \eqref{eq_siso_ynp1} is neglected, and when all actuators have the same bandwidth i.e. $\boldsymbol{\Omega}=\omega \boldsymbol{I}_{k \times k}$. For linear first-order actuators, Eq. \eqref{eq_siso_ynp1} can then be expressed by
\begin{equation}
    {{\boldsymbol{y}}^{(r + 1)}} = {{\boldsymbol{F}}_u}{\boldsymbol{\dot u}} = {{\boldsymbol{F}}_u}\omega \left( {{{\boldsymbol{u}}_c} - {\boldsymbol{u}}} \right)
    \label{eq_Fuudot}
\end{equation}
where the actuator relation in Eq. \eqref{eq_1storder_act} was substituted. Inserting the INDI control law from Eq. \eqref{eq_INDI_control_law_classic} into Eq. \eqref{eq_Fuudot} results in
\begin{equation}
    \boldsymbol{y}^{(r+1)}=\omega (\boldsymbol{\nu}-\boldsymbol{y}^{(r)}).
    \label{eq_nu_dy_assumption}
\end{equation}
i.e. $\boldsymbol{y}^{(r)}$ follows $\boldsymbol{\nu}$ with the dynamics of the actuator.
Based on Eq. \eqref{eq_nu_dy_assumption} a new control law can be derived. First the virtual control is defined by
\begin{equation}
    \boldsymbol{\nu}  = \frac{1}{\omega }{\boldsymbol{y}^{(r + 1)}} + {\boldsymbol{y}^{(r)}}
    \label{eq_act_ref}
\end{equation}
Choosing the same error dynamics like in the NDI control law, given by Eq. \eqref{eq_errordyn_exact} with $\boldsymbol{\Omega_y}=\omega \boldsymbol{I}_{m \times m}$, solving for $\boldsymbol{y}^{(r+1)}$, substituting this relation for $\boldsymbol{y}^{(r+1)}$ in Eq. \eqref{eq_act_ref}, and inserting the resulting $\boldsymbol{\nu}$ into the INDI control law given by Eq.  \eqref{eq_INDI_control_law_classic}, results in the control law
\begin{equation}
\begin{split}
        \boldsymbol{u_c} = \boldsymbol{F_u}^\dag \frac{1}{\omega }\Biggl( {\boldsymbol{y}_{ref}}^{(r + 1)} + \sum\limits_{i = 1}^r {{\boldsymbol{K}_{i-1}}\boldsymbol{e_y}^{(i)}}  
        + \omega \left( {\boldsymbol{e_y}^{(r)} + \sum\limits_{i = 0}^{r - 1} {{\boldsymbol{K}_i}\boldsymbol{e_y}^{(i)}} } \right) \Biggr) + {\boldsymbol{u}_0}
\end{split}
    \label{eq_ewoud_cl}
\end{equation}
This control law equals the NDI control law in Eq. \eqref{eq_siso_ctrl_law_final_complex}, if $\boldsymbol{\Omega} =\omega \boldsymbol{I}_{k \times k}$, $\boldsymbol{\Omega_y} = \omega \boldsymbol{I}_{m \times m}$ and $\boldsymbol{F_x} \boldsymbol{\dot{x}}$ is neglected. This approach requires all actuators to have the same bandwidth, otherwise Eq. \eqref{eq_nu_dy_assumption} does not hold.

In \cite{Raab2019} an approach can be found to resolve this problem. With the error dynamics proposed in Eq. \eqref{eq_errordyn_exact}, the control law proposed in \cite{Raab2019} can be formulated using the notation of this paper as  

\begin{equation}
\begin{split}
    \boldsymbol{u_c} = {(\boldsymbol{F_u}\boldsymbol{\Omega} )^\dag }\Biggl( {\boldsymbol{y}_{ref}}^{(r + 1)} + \sum\limits_{i = 1}^r {{\boldsymbol{K}_{i-1}}\boldsymbol{e_y}^{(i)}} 
    + \boldsymbol{\Omega _y}\left( {\boldsymbol{e_y}^{(r)} + \sum\limits_{i = 0}^{r - 1} {{\boldsymbol{K}_i}\boldsymbol{e_y}^{(i)}} } \right) \Biggr) + \boldsymbol{u}
\end{split}
\end{equation}
This is again equal to the corresponding NDI law in Eq. \eqref{eq_siso_ctrl_law_final_complex} with $\boldsymbol{F_x} \boldsymbol{\dot{x}}$ neglected.
Compared to Eq. \eqref{eq_ewoud_cl}, it does not have the assumption of equal actuators and it offers the option to specify a desired $\boldsymbol{\Omega_y}$.




\input{RollDynamicsExample}

\section{Discussion}\label{sec_discussion}

One of the potential issues that could be raised with the proposed sensor based NDI control law as given in Eq. \eqref{eq_siso_ctrl_law}, is the reliance on additional state information.
The question is if the respective signals can be measured or otherwise obtained.
This is something that is dependent on the system under consideration.
In terms of the system output, Eq. \eqref{eq:comparison} shows that the requirement is the same as for an INDI controller, the output up until $\boldsymbol{y}^{(r)}$ should be available.
Additionally, the \ac{NDI} control law requires the state derivative information $\boldsymbol{\dot{x}}$.
Depending on the system, this may overlap with the derivative of the output.
In many cases, the requirements on available signals may therefore be the same, or slightly increased compared to a regular \ac{INDI} controller.

In some cases, filtering of these noisy derivative signals may be required.
Filtering leads to delay, which will lead to deterioration of the controller performance.
For an \ac{INDI} controller, this problem can be circumvented with filter synchronization of output and input filters \cite{smeur2016}, but that particular method cannot be applied to the feedback of $\boldsymbol{\dot x}$ in the proposed controller, as there is no signal to synchronize with.
Instead, a complementary filter could be used, where the high frequencies are coming from a model, and the low frequencies from the filtered measurement \citep{Kim2021,Kumtepe2022}.
If the system under consideration is already stable, the proposed method could improve the tracking behavior, but the feedback of the state derivatives could influence the stability and robustness in the case of model uncertainty and measurement errors.
How this compares to classical INDI has to be compared on a case to case basis and should be further investigated in future research.
The proposed NDI control law additionally requires information about the actuator bandwidth, which might also be uncertain. In addition, the actuators might not be first-order. In that case, an effective bandwidth of the actuator can be used for the control law design.

\acresetall
\section{Conclusion}\label{sec_conclusion}
The proposed derivation of the \ac{NDI} control law considering first-order linear actuator dynamics, which one could call \ac{ANDI}, theoretically provides a perfect inversion of a system with such actuators, also in case the actuators do not all have the same bandwidth.
It allows to assign an innermost desired bandwidth for the tracking variable different than the actuator bandwidth.
The \ac{ANDI} control law leads to an \ac{INDI} control law if infinitely fast actuators are considered.
This provides a better theoretical foundation for INDI control, from which it can be readily seen what the impact is of the terms that are neglected in the \ac{INDI} derivation.
Moreover, the derived \ac{NDI} control law compensates for state-dependent terms, which are not taken into account with the classical \ac{INDI} formulation.
Compared to the reference model based feed-forward control from literature, the benefit of the proposed control law is twofold.
First, the inversion is based on the actual vehicle states, and therefore provides exact tracking performance without the reference model having to exactly model and match all plant states.
Secondly, it provides predictable and consistent error dynamics.

\section{Acknowledgements}
The authors would like to thank Stefan Raab and Florian Holzapfel for the fruitful discussions of INDI concept development and their many interesting insights into the theory behind incremental control laws.
The authors would also like to thank the reviewers for their qualified comments and suggestions.


\input{Appendix_limits}




\bibliography{references}

\end{document}

%% file: RollDynamicsExample.tex
\section{Simple Example}\label{sec_example}
In the following, the mechanics of deriving the control law is demonstrated on a simple single-input single-output (SISO) linear system. The roll rate of a fixed wing aircraft is to be controlled. Consider the roll dynamics given by:
\begin{equation}
    \dot p = {L_p}p + {L_u}u
    \label{eq_roll_dyn}
\end{equation}
where $p$ is the roll rate, $L_p<0$
is the roll damping, u is the aileron deflection, and $L_u<0$ is the aileron roll effectiveness. $L_p$ is usually large, and hence contributes significantly to the dynamics. This example demonstrates how it is taken into account by the proposed control concept. The ailerons are driven by an actuator with the following first-order dynamics:
\begin{equation}
    \dot u = \omega \left( {{u_c} - u} \right)
\end{equation}
where $u_c$ is the actuator command and $\omega$ is the actuator bandwidth. The output to be controlled is $y = p$. The relative degree $r$ of the system is $1$ as given by Eq. \eqref{eq_roll_dyn}. The control relation as given by Eq. \eqref{eq_siso_ypp1} is:
\begin{equation}
    \ddot p = {L_p}\dot p + {L_u}\omega \left( {{u_c} - u} \right)
\end{equation}
Solving for $u_c$ as explained in Eq. \eqref{eq_siso_ctrl_law}, gives the following control law.
\begin{equation}
    {u_c} = \frac{1}{{{L_u}\omega }}\left( {{{\ddot p}_d} - {L_p}\dot p} \right) + u
    \label{eq_ctrl_law_ex}
\end{equation}
A reference model is chosen based on the desired dynamics:
\begin{equation}
\begin{split}
{{\dot p}_{ref}} & = {-L_{p,d}}\left( { {\delta} - {p_{ref}}} \right)\\
\dot \delta &= {\omega _d}\left( {{p_c} - {\delta}} \right)
\end{split}
\label{eq_reference_dynamics}
\end{equation}
where $L_{p,d}<0$ is the desired roll damping and $\delta$ 
is a generalized roll acceleration due to the aileron deflection.

\begin{figure}[ht]
	\centering
	\includegraphics[scale=0.8]{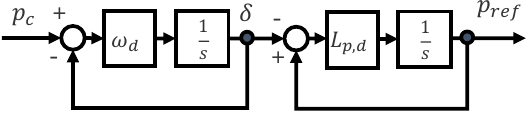}
	\caption{Block diagram of the reference model}  
	\label{fig_BD} 
\end{figure}

The dynamics of $\delta$ is given by the desired build up in roll acceleration due to the aileron dynamics $\omega_d$. ${{p_c}}$ is the pilot roll rate command, since in steady state, ${p_{ref}} = {p_c}$. Note that ${{\ddot p}_{ref}}$ can easily be calculated using the reference dynamics in Eq. \eqref{eq_reference_dynamics}:
\begin{equation}
{{\ddot p}_{ref}} =  - {L_{p,d}}\left( {{\omega _d}\left( {{p_c} - \delta } \right) - {{\dot p}_{ref}}} \right)
\label{eq_NDI_ref_example2}
\end{equation}
To choose ${\ddot p}_d$, the desired error dynamics have to be formulated. Since the actuators are of first-order, and the system dynamics are first-order, a product of 2 first-order systems are chosen as follows:
\begin{equation}
    {E_p}\left( {s + {\omega _d}} \right)\left( {s - {L_{p,d}}} \right) =  0
\end{equation}
where $e_p = p_{ref}-p$ and which in the time domain is:
\begin{equation}
{{\ddot e}_p} + {\omega _d}{{\dot e}_p} - {L_{p,d}}\left( {{{\dot e}_p} + {\omega _d}{e_p}} \right) = {{\ddot e}_p} +k_1{\dot e}_p + k_0{e_p}  = 0
\label{eq_error_dyn_des_NDI}
\end{equation}
where 
\begin{equation}
\begin{array}{l}
{k_1} = \left( {{\omega _d} - {L_{p,d}}} \right)\\
{k_0} =  - {L_{p,d}}{\omega _d}
\end{array}
\end{equation}
Solving for $\ddot p = \ddot p_d$:
\begin{equation}
    {{\ddot p}_d} = {\ddot p}_{ref} + k_1 \dot{e}_p +k_0 {e}_p 
    \label{eq_pddot_d}
\end{equation}
will lead to the final control law by substituting Eq. \eqref{eq_pddot_d} into Eq. \eqref{eq_ctrl_law_ex}:
\begin{equation}
    {u_c} = \frac{1}{{{L_u}\omega }}\left( {\ddot p}_{ref} + k_1 \dot{e}_p +k_0 {e}_p - {L_p}\dot p \right) + u
    \label{eq_ndi_ex_uc}
\end{equation}
which can be ordered in terms of the contributions:
\begin{equation}
{u_c} = \frac{1}{{{L_u}\omega }}\left( {\underbrace {{{\ddot p}_{ref}}}_{{\text{Feed-forward}}} + \underbrace {{k_1}{{\dot e}_p} + {k_0}{e_p}}_{{\text{Error control}}} - \underbrace {{L_p}\dot p}_{{\text{Model part}}}} \right) + u
\label{eq_NDI_example}
\end{equation}
This can be compared to the standard INDI control law \eqref{eq:indi_with_dx} for the same example 
\begin{equation}
{u_c} = \frac{1}{{{L_u}}}\left( \dot{e}_p- {L_{p,d}}{e_p} \right) + u
\label{eq_INDI_example}
\end{equation}
with reference dynamics:
\begin{equation}
    {{\dot p}_{ref}} =  - {L_{p,d}}\left( {{p_c} - {p_{ref}}} \right)
    \label{eq_INDI_example_ref}
\end{equation}

\subsection{Simulation results}
For the example described above, simulations were conducted to validate the approach.
In the simulations, it was assumed that state information was available without noise, and there were no uncertainties in the parameters. The parameters were $L_u=0.25$ (deg/s$^2$)/deg, $L_p = -6.6$ (deg/s$^2$)/(deg/s),  $L_{p,d} = 2L_p = -13.2$ (deg/s$^2$)/(deg/s) and $\omega=20$ rad/s.
In the following we will compare:
\begin{enumerate}
    \item \textit{NDI} as given in Section \ref{sec:ndi}, for this example given by Eq. \eqref{eq_ndi_ex_uc}, with reference dynamics given by Eq. \eqref{eq_reference_dynamics} and Eq. \eqref{eq_NDI_ref_example2}.  
    \item \textit{INDI} as given in Section \ref{sec:indi}, for this example given by Equation \eqref{eq_INDI_example} with reference dynamics given by \eqref{eq_INDI_example_ref}.
    \item \textit{INDI with actuators} as given in Section \ref{sec_INDI_w_actuators}, for this example given by Eq. \eqref{eq_ndi_ex_uc} without $L_p \dot p$, with reference dynamics given by Eq. \eqref{eq_reference_dynamics} and Eq. \eqref{eq_NDI_ref_example2}.
\end{enumerate}


\begin{figure}[ht]
  \centering
  \includegraphics[scale=0.6]{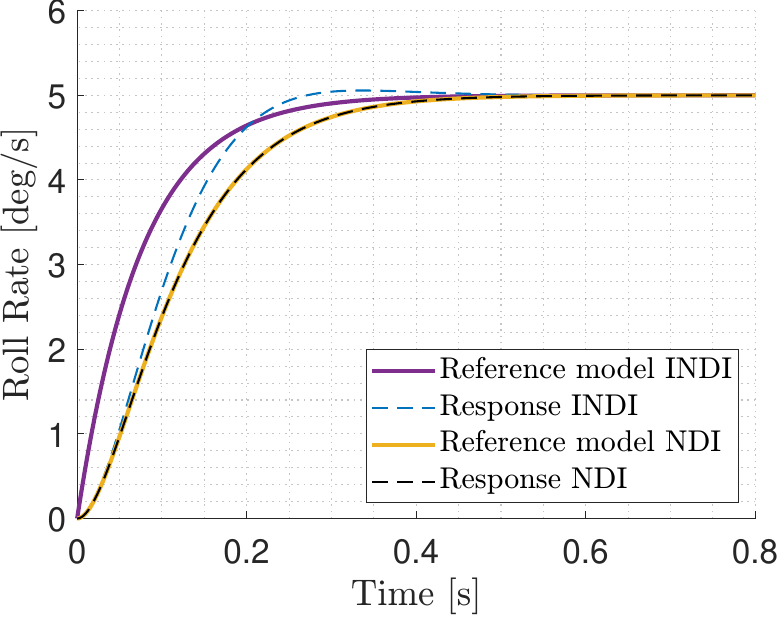}
  \caption{Step response of the INDI and NDI control law.}
\label{fig_Step_l}
\end{figure}
\begin{figure}[ht]
  \centering
  \includegraphics[scale=0.6]{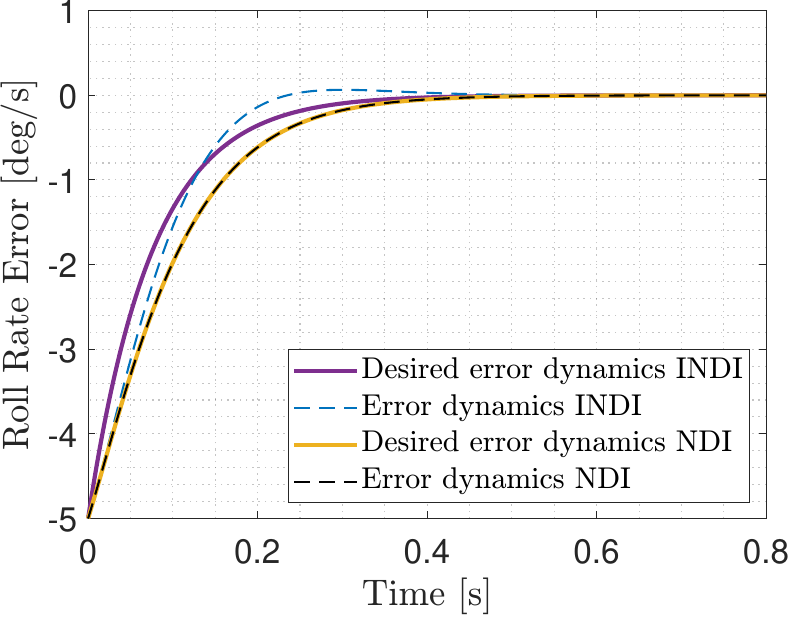}
   \caption{Comparison of desired and actual error dynamics considering a $5^\circ/s$ initial value perturbation.}
  \label{fig_Step_r}
\end{figure}

\begin{figure}[ht]
	\centering
    \includegraphics[scale=0.6]{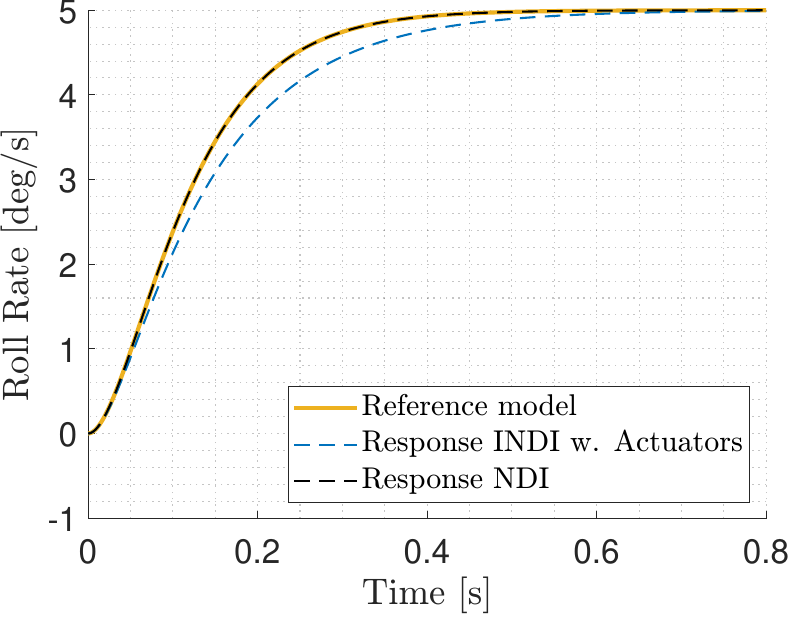}
	\caption{Step response of the INDI with Actuators in the design and NDI control law.}
	\label{fig_Step2_l} 
\end{figure}
\begin{figure}[ht]
      \centering
      \includegraphics[scale=0.6]{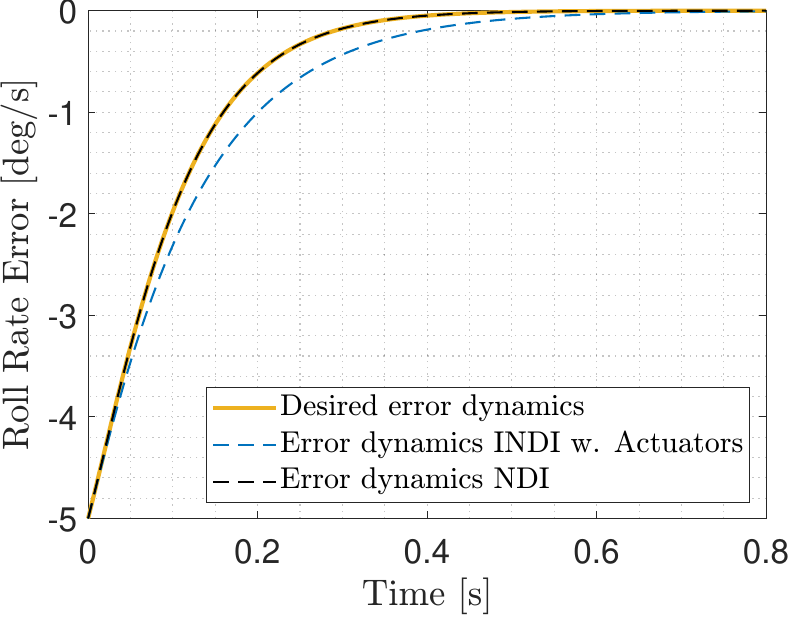}
	\caption{Comparison of desired and actual error dynamics considering a $5^\circ/s$ initial value perturbation.}
	\label{fig_Step2_r} 
\end{figure}

Figure \ref{fig_Step_l} compares the responses in roll rate $p$ to a step input of $5^\circ/s$ in $p_{c}$ for the INDI and NDI control law. 
The NDI law makes the plant correctly follow the reference model with reference inputs, while this is not the case for INDI with its corresponding reference dynamics.

Figure \ref{fig_Step_r} reveals the error dynamics of the classical INDI control law in Eq. \eqref{eq_INDI_example} and the proposed NDI law in Eq. \eqref{eq_NDI_example}.
The closed loop system response for both control laws was simulated with a roll rate command $p_c=0^\circ/s$ for an initial value perturbation of $p=5^\circ/s$, i.e. the initial value of the plant was $5^\circ/s$, while the initial value of the reference model was $p_{ref}=0^\circ/s$.
The simulation shows that the resulting error dynamics of the NDI controller correspond exactly to the desired error dynamics given by Eq. \eqref{eq_error_dyn_des_NDI}.
The resulting error dynamics of the INDI controller do not correspond to the specified error dynamics $\dot{e}_p-L_{p,d}e_p=0$.

Figure \ref{fig_Step2_l} compares the NDI control law with the INDI control law where actuator dynamics are taken into account in the error controller design.
Both laws use the same reference dynamics.
This extended INDI controller still does not realize perfect tracking of the reference signal of the NDI law, the reason being that in the INDI controller the $L_p \dot p$ term is neglected.
In \cite{Bhardwaj2021}, this issue was fixed for the extended INDI controller by adding an additional reference model based feed-forward term.
The resulting control law in \cite{Bhardwaj2021} can be reformulated to resemble the NDI control law given by Eq. \eqref{eq_NDI_example}, with the only difference that the term $L_p \dot p$ is substituted with $L_p \dot p_{ref}$ from a physical reference model, such as the one given in Eq. \eqref{eq_reference_dynamics}.
Under the condition that $\dot p=\dot p_{ref}$, i.e. that no error exist, and the plant is exactly equal to the reference model, this leads to perfect tracking of reference inputs, similar to the NDI law.
However, the condition that $F_x \dot x$ equals $F_x \dot x_{ref}$, might not hold in the following cases: 1) dynamics and couplings which are not modeled in the reference dynamics but are present in the plant, 2) a control allocation mismatch for the reference model and the INDI controller due to disturbances, model uncertainties, and utilization of over-actuated channels for secondary control objectives, leading to different responses, and 3) disturbances.

In Figure \ref{fig_Step2_r} the discrepancy is shown for the case that the reference model state is zero, but the plant state is not.
Here, the closed-loop system responses of the NDI control law and the INDI with actuators were simulated for a roll rate command $p_c=0^\circ/s$ with an initial value of the plant of $5^\circ/s$, while the initial value of the reference model was $p_{ref}=0^\circ/s$, such that $\dot p \neq \dot p_{ref}=0$.
It is revealed that for the NDI law in Eq. \eqref{eq_NDI_example}, the dynamics with which the error $p_{ref}-p$ declines, corresponds exactly to the desired error dynamics specified by Eq. \eqref{eq_error_dyn_des_NDI}. 
The INDI control law with actuators was designed with the same desired error dynamics like the NDI law and it is seen that the resulting error dynamics do not correspond to these desired dynamics.

\begin{figure}[ht]
	\centering
    \centering
    \includegraphics[scale=0.6]{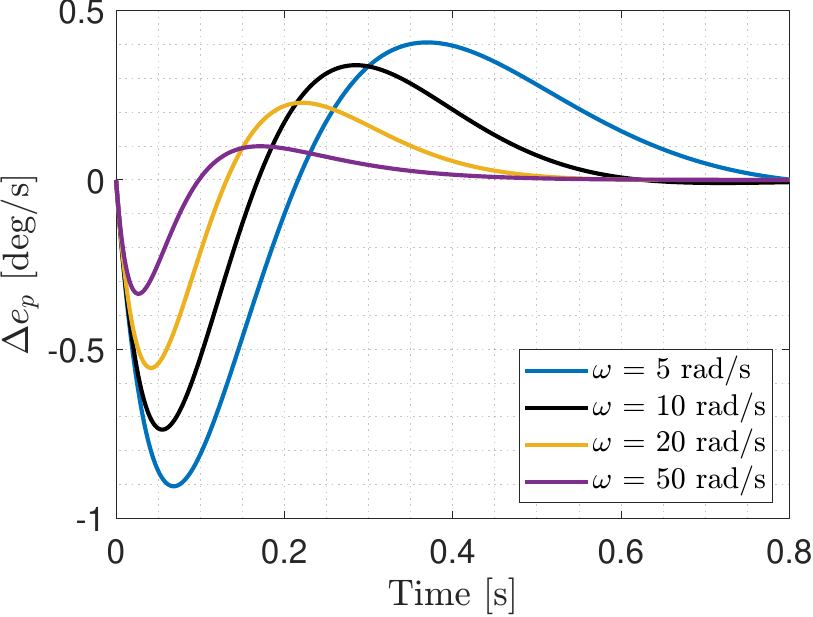}
	\caption{Difference between actual INDI error dynamics and design INDI error dynamics.}
	\label{fig_ED_INDI_INDI_l} 
\end{figure}

\begin{figure}[ht]
      \centering
      \includegraphics[scale=0.6]{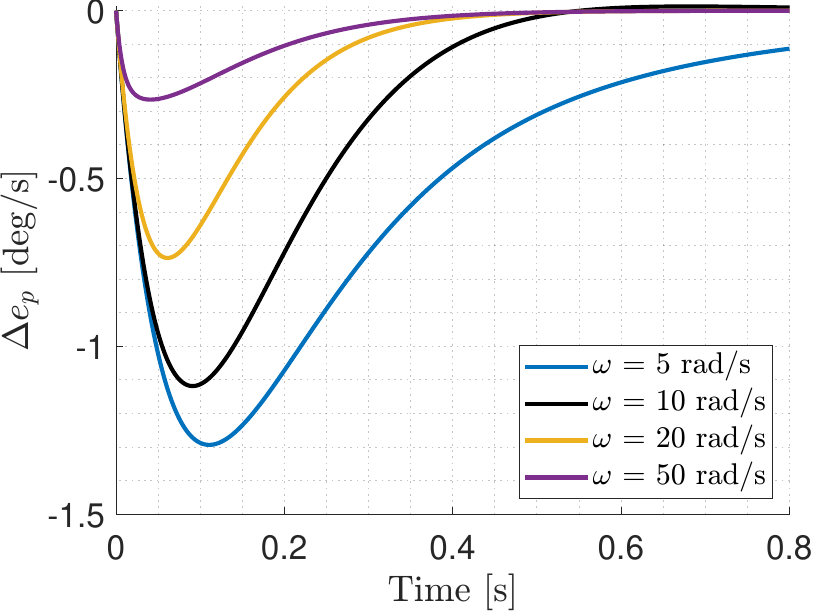}
	\caption{Difference between NDI error dynamics and design INDI error dynamics.}
	\label{fig_ED_INDI_INDI_r}
\end{figure}

Finally, Fig. \ref{fig_ED_INDI_INDI_l} and Fig. \ref{fig_ED_INDI_INDI_r} investigate the influence of the actuator bandwidth on the error dynamics of the classical INDI and the NDI control laws.
The same simulations as before with $p_c=0^\circ/s$ and an initial condition  $p=5^\circ/s$ was performed, but with varying actuator bandwidth.
The simulations show that:
 \begin{enumerate}
 \item For the INDI law in Eq. \eqref{eq_INDI_example} the dynamics of $p_{ref}-p$ do not exactly correspond to the error dynamics that were specified in the design, i.e. $\dot{e}-L_{p,d}e_p=0$, but for increasing actuator bandwidth these dynamics are approached by the resulting error dynamics.
 \item The desired NDI error dynamics given by Eq. \eqref{eq_error_dyn_des_NDI} approach the INDI error dynamics given by $\dot{e}-L_{p,d}e_p=0$, for increasing actuator bandwidth.
 Because the INDI error dynamics do not explicitly consider actuator dynamics, then when the actuator bandwidth increases, the actuator part of the error dynamics in the NDI law becomes negligible and hence approaches the INDI error dynamics.
 \end{enumerate}


Consider for Statement 1) Figure \ref{fig_ED_INDI_INDI_l}, which reveals the \textit{difference} between the error $e_{p}$ resulting from simulation of the closed-loop system with the INDI control law, and the initial value response of the desired error dynamics $\dot{e}-L_{p,d}e_p=0$. It shows that for increasing bandwidth of the actuator this difference decreases, meaning that the resulting error dynamics approach the specified desired dynamics of the INDI law. 
For Statement 2) consider Figure \ref{fig_ED_INDI_INDI_r}, which depicts the \textit{difference} between the desired INDI error dynamics and the desired NDI error dynamics, for increasing actuator bandwidth. Hence, the conclusion can be drawn that with increasing actuator bandwidth, the actuator part of the error dynamics in the NDI law becomes negligible, and the NDI error dynamics approach the INDI error dynamics.

%% file: Appendix_limits.tex
\section*{Appendix: NDI and INDI limit}
\label{appendix}
Here it is shown how the error dynamics of NDI and INDI behave in the limit of the actuator bandwidth going to infinity.
Consider first the NDI law from equation \eqref{eq_siso_ctrl_law2} together with the system output  dynamics as given in equation \eqref{eq_siso_ypp1}, with the same assumptions as in Section \ref{sec:indi}, i.e.
\begin{enumerate}
    \item All actuators have the same bandwidth, i.e. $\boldsymbol{\Omega}=\omega \boldsymbol{I}_{k\times k}$, where $\omega$ is a non-zero scalar
    \item Choose $\boldsymbol{\Omega_y}=\omega \boldsymbol{I}_{m\times m}$
\end{enumerate}
By inserting Eq. \eqref{eq_siso_ctrl_law2} into the system dynamics given in Eq. \eqref{eq_siso_ypp1}, the error dynamics can be deduced:
\begin{equation}
\begin{array}{*{20}{l}}
  {{{\boldsymbol{y}}^{(r + 1)}} = {{\boldsymbol{F}}_{\boldsymbol{x}}}{\boldsymbol{\dot x}} + {{\boldsymbol{F}}_{\boldsymbol{u}}}{{\boldsymbol{I}}_{k \times k}}\omega \left( {{{\boldsymbol{u}}_{\boldsymbol{c}}} - {\boldsymbol{u}}} \right)} \\ 
  {{{\boldsymbol{y}}^{(r + 1)}} = {{\boldsymbol{F}}_{\boldsymbol{x}}}{\boldsymbol{\dot x}} + {{\boldsymbol{F}}_{\boldsymbol{u}}}{{\boldsymbol{I}}_{k \times k}}\omega \left( {\frac{1}{\omega }{{({{\boldsymbol{F}}_{\boldsymbol{u}}})}^ + }\left( {{\boldsymbol{y}}_{ref}^{(r + 1)} + \sum\limits_{i = 1}^r {{{\boldsymbol{K}}_{i - 1}}} {{\boldsymbol{e}}_{\boldsymbol{y}}}^{(i)} - {{\boldsymbol{F}}_{\boldsymbol{x}}}{\boldsymbol{\dot x}}} \right) + {{({{\boldsymbol{F}}_{\boldsymbol{u}}})}^ + }\left( {{{\boldsymbol{e}}_{\boldsymbol{y}}}^{(r)} + \sum\limits_{i = 0}^{r - 1} {{{\boldsymbol{K}}_i}{{\boldsymbol{e}}_{\boldsymbol{y}}}^{(i)}} } \right) + {\boldsymbol{u}} - {\boldsymbol{u}}} \right)} \\ 
  {{{\boldsymbol{y}}^{(r + 1)}} = {{\boldsymbol{F}}_{\boldsymbol{x}}}{\boldsymbol{\dot x}} + \frac{1}{\omega }\omega \left( {{\boldsymbol{y}}_{ref}^{(r + 1)} + \sum\limits_{i = 1}^r {{{\boldsymbol{K}}_{i - 1}}} {{\boldsymbol{e}}_{\boldsymbol{y}}}^{(i)} - {{\boldsymbol{F}}_{\boldsymbol{x}}}{\boldsymbol{\dot x}}} \right) + \omega \left( {{{\boldsymbol{e}}_{\boldsymbol{y}}}^{(r)} + \sum\limits_{i = 0}^{r - 1} {{{\boldsymbol{K}}_i}{{\boldsymbol{e}}_{\boldsymbol{y}}}^{(i)}} } \right)} \\ 
  {{{\boldsymbol{y}}^{(r + 1)}} = \left( {{\boldsymbol{y}}_{ref}^{(r + 1)} + \sum\limits_{i = 1}^r {{{\boldsymbol{K}}_{i - 1}}} {{\boldsymbol{e}}_{\boldsymbol{y}}}^{(i)}} \right) + \omega \left( {{{\boldsymbol{e}}_{\boldsymbol{y}}}^{(r)} + \sum\limits_{i = 0}^{r - 1} {{{\boldsymbol{K}}_i}{{\boldsymbol{e}}_{\boldsymbol{y}}}^{(i)}} } \right)} \\ 
  {{{\boldsymbol{e}}_{\boldsymbol{y}}}^{(r + 1)} + \sum\limits_{i = 1}^r {{{\boldsymbol{K}}_{i - 1}}} {{\boldsymbol{e}}_{\boldsymbol{y}}}^{(i)} + \omega \left( {{{\boldsymbol{e}}_{\boldsymbol{y}}}^{(r)} + \sum\limits_{i = 0}^{r - 1} {{{\boldsymbol{K}}_i}{{\boldsymbol{e}}_{\boldsymbol{y}}}^{(i)}} } \right) = {\boldsymbol{0}}} 
\end{array}
\end{equation}
which is the desired error dynamics given in Eq.  \eqref{eq_td_dyn}.
\\
If the INDI law from \eqref{eq:comparison} is inserted into Eq. \eqref{eq_siso_ypp1}, the following can be derived:
\begin{equation}
\begin{array}{*{20}{l}}
  {{{\boldsymbol{y}}^{(r + 1)}} = {{\boldsymbol{F}}_{\boldsymbol{x}}}{\boldsymbol{\dot x}} + {{\boldsymbol{F}}_{\boldsymbol{u}}}{{\boldsymbol{I}}_{k \times k}}\omega \left( {{{({{\boldsymbol{F}}_{\boldsymbol{u}}})}^\dag }\left( {{{\boldsymbol{e}}_{\boldsymbol{y}}}^{(r)} + \sum\limits_{i = 0}^{r - 1} {{{\boldsymbol{K}}_i}{{\boldsymbol{e}}_{\boldsymbol{y}}}^{(i)}} } \right) + {\boldsymbol{u}} - {\boldsymbol{u}}} \right)} \\ 
  {{{\boldsymbol{y}}^{(r + 1)}} = {{\boldsymbol{F}}_{\boldsymbol{x}}}{\boldsymbol{\dot x}} + {{\boldsymbol{F}}_{\boldsymbol{u}}}{{\boldsymbol{I}}_{k \times k}}\omega \left( {{{({{\boldsymbol{F}}_{\boldsymbol{u}}})}^\dag }\left( {{{\boldsymbol{e}}_{\boldsymbol{y}}}^{(r)} + \sum\limits_{i = 0}^{r - 1} {{{\boldsymbol{K}}_i}{{\boldsymbol{e}}_{\boldsymbol{y}}}^{(i)}} } \right)} \right)} \\ 
  { - {{\boldsymbol{y}}^{(r + 1)}} + {{\boldsymbol{F}}_{\boldsymbol{x}}}{\boldsymbol{\dot x}} + \omega \left( {{{\boldsymbol{e}}_{\boldsymbol{y}}}^{(r)} + \sum\limits_{i = 0}^{r - 1} {{{\boldsymbol{K}}_i}{{\boldsymbol{e}}_{\boldsymbol{y}}}^{(i)}} } \right) = {\boldsymbol{0}}} 
\end{array}
	\label{eq:indilimit}
\end{equation}
When taking the limit of $\omega$ tending to infinity, Eq. \eqref{eq:indilimit} does not tend to the desired error dynamics.
Hence, even in the limit, the INDI does not produce the correct error dynamics.
INDI only approximates the true NDI control law signal $\boldsymbol{u}_c$ arbitrarily well as given in Eq. \eqref{eq:comparison}.